# Crystal Nucleation Modeling of Solvent Molecules Influence on Radius and Morphology of Nano Copper Ferrite Particles

Zihan Huang[*1], Yiyi Hang[2]

1. College of Chemistry and Molecular Engineering
   Peking University (PKU), Beijing 100871, China
   25/Nov/2021

2. Department of Material Science and Engineering
   Harbin Engineering University (HEU)
   Harbin 150001, CN

**Abstract**:

   Nanometer copper ferrite, as a kind of nanometer particle with catalytic activity, and its photothermal and magnetothermal effects as ferrite, can be widely used in different fields. It is a general way to obtain the nano effect of the target by controlling the particle size. In this paper, the crystallization process of hydrothermal/solvothermal synthesis was analyzed, and the nucleation model was established to simulate the effects of solvent, reaction temperature and cooling time on the particle size of copper ferrite nanoparticles. Through Monte Carlo method and energy function, the ratio of nano particle agglomeration was established, and the influence of different reaction conditions on it was discussed.

# Introduction

With the development of nano science, more and more researches have been done on nano effects. At the same time, researchers are also looking for methods to control particle nano effects. The nano effect mainly comes from the nano scale of particles, at which the physical and chemical properties of substances can be greatly changed. The size of nanoparticles is directly related to their size, which depends on their crystallization process. By changing the experimental conditions, such as solvent, temperature, pressure, reaction time, etc., the crystallization process of substances can be affected [1], and then different particle sizes can be obtained.

Copper ferrite has catalytic activity for hydrogen peroxide in nanometer scale, and different particle sizes have great influence on its catalytic activity [2]. Therefore, controlling the particle size during the synthesis process is a widely used method. Then, by simulating and analysing the influence of factors on particle size in synthesis, guidance for its preparation process can be provided. The hydrothermal/ solvothermal synthesis is a general synthesis method used when preparing copper ferrite [3][4][5], as it is low cost and easy to operate. In this article, the modeling of nucleation process will base on hydrothermal/ solvothermal synthesis.

# 1 Model Design

**1.1 Energy Change during Nucleus Formation**

The modeling focus on the influence of solvents on the size of nanoparticles produced. The crystal nucleus are formed to reduce the Gibbs freedom energy of the system [6]. The change of Gibbs freedom energy from ions in solvents to crystal cell structure per unit volume was written as $\Delta G_V$. There is following chemical equation to express the crystallization.

$$2Fe^{3+} + Cu^{2+} + 8OH^- \rightarrow CuFe_2O_4 + 4H_2O$$

The standard Gibbs energy of molar formation of $CuFe_2O_4$ can be calculated by the above equation, which was expressed as formulation (F.1).

$$\Delta_f G_m^\ominus\ CuFe_2O_4 = 2\Delta_f G_m^\ominus\ Fe^{3+} + \Delta_f G_m^\ominus\ Cu^{2+} \\ + 8\Delta_f G_m^\ominus\ OH^- - 4\Delta_f G_m^\ominus\ H_2O \quad (F.1)$$

$\Delta_f G_m^\ominus\ Fe^{3+} = -4.7 kJ/mol$
$\Delta_f G_m^\ominus\ Cu^{2+} = 65.49 kJ/mol$
$8\Delta_f G_m^\ominus\ OH^- = -157.2 kJ/mol$
$\Delta_f G_m^\ominus\ H_2O = -237.1 kJ/mol$

The alpha lattice parameter $\alpha$ of $CuFe_2O_4$ is 8.389(2) Å [7]. The unit cell of the spinel structure can be expressed as $Cu_8Fe_{16}O_{32}$, the 32 oxygen ions are faced center cubic (FCC). Each cubic cell is consisted by eight $CuFe_2O_4$ units [8]. Thus,

$$\Delta G_V = \frac{8}{\alpha^3}\Delta_f G_m^\ominus\ CuFe_2O_4 \quad (F.2)$$

During the process of crystal nucleus uniform formation, when a sphere embryonic crystal with radius $r$ exists in solvent, the total change of Gibbs freedom energy is

$$\Delta G = \frac{4}{3}\pi r^3 \Delta G_V + 4\pi r^2 \sigma \quad (F.3)$$

where the $\sigma$ is specific surface energy. The critical radius $r^*$ is obtained by applying $\frac{d\Delta G}{dr} = 0$ on (F.3),

$$r^* = -\frac{2\sigma}{\Delta G_V} \quad (F.4)$$

Thus, by applying formulation (F.4) in (F.3), the work needed to form a critical crystal nucleus can be expressed by

$$\Delta G^* = \frac{16\pi\sigma^3}{3(\Delta G_V)^2} \quad (F.5)$$

To form the crystal nucleus, $\Delta G^*$ need to be filled with the energy fluctuation in solvents, which was considered as the summation of kinetic energy $E_k$ of all solvents molecules the nucleus contacted during its formation. Thus, the amount of solvent molecules contacted is the amount of solvent molecules in volume same as the volume of crystal nucleus. $E_k$ is:

$$E_k = \frac{3}{2}k_B T \cdot \frac{4\pi r^{*3} \cdot \rho_s}{3M_s} \cdot N_A \quad (F.6)$$

The section before the first multiplication sign is the average kinetic of solvent molecules. $k_B$ is the Boltzmann constant, which is $1.381 \times 10^{-23}$ m$^2$ kg s$^{-2}$ /K, $T$ is the absolute temperature of solvent, $\rho_s$ is the density of solvent, $M_s$ is the molar mass of solvent, $N_A$ is Avogadro constant, which is $6.022 \times 10^{23}$ mol$^{-1}$.

At the critical situation, $\sigma$ at this situation can be solved by applying formulation (F.6) equal to (F.5). Then $\Delta G^*$ can be solved by (F.5).

## 1.2 Nucleation Rate

The nucleation rate $N$ (m$^{-3}$ s$^{-1}$) is the amount of crystal nucleus formed by a unit volume solvent in a unit time [9], it was controlled by two coefficients. In formulation (F.7), the one before the multiplication sign is coefficient of work of nucleation, the other one is coefficient of diffusion probability of atoms.

$$N = K\exp\left(\frac{-\Delta G^*}{k_B T}\right) \cdot \exp\left(\frac{-Q}{k_B T}\right) \quad (F.7)$$

where $K$ is proportionality constant, $Q$ (J/mol) is energy of diffusion activity of atoms, which is related to atomic binding force as well as mechanism of diffusion. It can be calculated by introducing P-parameters with effective energy of paired interaction of atoms, as it was shown in (F.8) [10].

$$\frac{1}{E_b} = 2\left[\left(\frac{r_i n}{P_0}\right)_1 + \left(\frac{r_i n}{P_0}\right)_2\right] \quad (F.8)$$

where $E_b$ is same meaning as the $Q$, but under eV unit. $r_i$ is the orbital radius of i–orbital of the atom, $n$ is the number of effective valence electrons, $P_0$ was called a spatial-energy parameter. For $CuFe_2O_4$, it was considered that Cu atoms take the position of crystal cell first then the iron and oxygen atoms diffused in. The parameters needed for calculation were listed in Table.1 [10].

**Table.1** $P_0$-parameters of valence orbitals of neutral atoms in basic state [10]

| Atom | Valence orbitals | $r_i$(Å) | $P_0$ (eVÅ) |
|---|---|---|---|
| O | $2p^1$ | 0.414 | 5.225 |
|   | $2p^1$ | 0.414 | 12.079 |
| Fe(III) | $3d^1$ | 0.365 | 10.564 |
| Cu(II) | $3d^1$ | 0.312 | 6.191 |

### 1.3 Temperature Range

In most preparation, after keeping a higher reaction temperature, there is a cooling process over the system that contains product [11][12]. During this process, there is a maximum and a minimum temperature, $T_{max}$, $T_{min}$, for the crystal nucleus forming and growing.

If $T_{max}$ is above the reaction temperature $T_R$, then $T_{max}$ was equaled to $T_R$. If not, over the maximum temperature, the Gibbs freedom energy of nucleation reaction is >0, which cannot process spontaneously. Thus, by applying $\Delta_f G_m^\ominus$ $CuFe_2O_4$ in (F.1) as a function of temperature (F.9), then equal it to 0, $T_{max}$ can be solved. Although the Gibbs freedom energy of formation was seen as a constant previous, the model need a border condition for calculation, which has to consider that at here it is related to temperature.

$$\Delta_f G_m^\ominus = \Delta_f H_m^\ominus - T_{max}\Delta_f S_m^\ominus = 0 \qquad (F.9)$$

The crystal is not able to growth below $T_{min}$, which means that the atoms in solvent cannot get through in crystal. Thus, the $T_{min}$ can be solved by a critical situation, when the kinetic energy of 1mol atoms in solvent is equal to the energy of diffusion activity of the 1mol atoms to diffuse into the crystal structure. The formulation can be written as (F.10).

$$T_{min} = \frac{2Q}{3k_B \cdot N_A} \qquad (F.10)$$

Consider the cooling as a linear process, $\beta$ is lapse rate of temperature, $t$ is time (s) since the temperature reaches $T_{max}$. As the experiments systems cool down from 280 ℃ to 60 ℃ in 60min, so $\beta$ was set as 0.0583K/s. There is

$$T = T_{max} - 0.0583\, t \qquad (F.11)$$

### 1.4 Average Radius of Nanoparticles

The total amount of crystal nucleus formed at time $t$ since the temperature reaches $T_{max}$ is $n_c$, it is the summation of the nucleation rate $N$, which is a function of $T$, from $T_{max}$ to $T_{min}$ in the whole volume of solvent $V_s$. Thus,

$$n_c = \int_0^t N[T(t)] \cdot V_s dt \qquad (F.12)$$

The $V_s$ was set as 30mL in experiments [13]. When it cools down to $T_{min}$, the total time cost can be calculated by (F.11), which is ($T_{max}$-$T_{min}$)/0.0583.

Assume the copper is excess in reaction, there are formulation (F.13a) based on iron conservation. The amount of Fe atoms in crystal equal to that in reactants.

$$16 \times \frac{4\pi R^3}{3\alpha^3} \cdot n_c = n_D \cdot N_A \xrightarrow{t=\infty} n_{Fe} \cdot N_A \qquad (F.13a)$$

where $R$ is the average radius of product nanoparticles, $n_D$ is the mole of iron diffused in nucleus, when $t=\infty$, because of the copper excess, all iron was considered to form product, $n_{Fe}$ is the mole of iron in reactants, which was set as 1mol in experiments. $n_D$ is denoted by (F.13b). J is diffusive mass transfer flux of Fe ions.

$$n_D = \int_0^t J \cdot n_c \cdot 4\pi R^2\, dt \qquad (F.13b)$$

### 1.5 Morphology of Nanoparticles Aggregation

The aggregation of nano particles influence the morphology seriously. As the aggregation can reduce the specific surface energy of nanoparticles. However, the particles have to overcome repulsion between each other during the aggregation process. To simulate the aggregation process, Monte Carlo method was applied.

The hexagonal center packing (HCP) has the highest density of packing modes, which is 74% [14]. It is the most efficient way to reduce specific surface energy of the particles aggregation. In HCP, each ball is surrounded by 12 other balls (Fig.1). Choose one part $L_0$ with vector $\vec{R}$ of its location, which is a single sphere particle initial at origin, the next tangent particle $L_1$ aggregates from one direction of the 12 positions randomly. The way is to choose one random number $p_1$ generated in (0, 12], and then create a random displacement vector $\vec{\Xi}$ in the sphere reference frame.

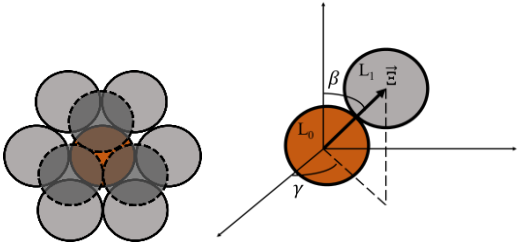

Fig.1 Structure of HCP and vector representation in sphere reference frame

If $p_1$ is in (0, 1], $\beta=\pi/2$, $\gamma=0$. If $p_1$ is in (1, 2], $\beta=\pi/2$, $\gamma=0$. If $p_1$ is in (2, 3], $\beta=\pi/2$, $\gamma=2\pi/3$. If $p_1$ is in (3, 4], $\beta=\pi/2$, $\gamma=\pi$. If $p_1$ is in (4, 5], $\beta=\pi/2$, $\gamma=4\pi/3$. If $p_1$ is in (5, 6], $\beta=\pi/2$, $\gamma=5\pi/3$. If $p_1$ is in (6, 7], $\beta=\pi/6$, $\gamma=\pi/6$. If $p_1$ is in (7, 8], $\beta=\pi/6$, $\gamma=5\pi/6$. If $p_1$ is in (8, 9], $\beta=\pi/6$, $\gamma=3\pi/2$. If $p_1$ is in (9, 10], $\beta=5\pi/6$, $\gamma=\pi/6$. If $p_1$ is in (10, 11], $\beta=5\pi/6$, $\gamma=5\pi/6$. If $p_1$ is in (11, 12], $\beta=5\pi/6$, $\gamma=3\pi/2$. With $R$ is the average radius of product nanoparticles calculated in section 1.4, the module of $\vec{\Xi}$ is

$$\|\vec{\Xi}\| = 2R \qquad (F.14)$$
$$\vec{\Xi} = [\beta, \gamma, 2R]$$

There is a restriction that the position of the new particle $L_1$ must not coincide with the existing particles in $L_0$. For an example, if 4 of 12 positions of $L_0$ were already placed, $L_1$ can only be random placed in the remainder 8 positions.

The energy difference during this process is $\delta\varepsilon$, which is expressed by the difference between specific surface energy change $\Delta E_S$ to Lennard-Jones potential $V_{L-J}$.

$$\delta\varepsilon = |\Delta E_S| - V_{L-J} \qquad (F.15)$$

Check whether $\delta\varepsilon$ is positive or negative, if e is positive, replace $\vec{R}$ by $\vec{R}+\vec{\Xi}$ and return to next particle. If e is negative, pick a new random number k in [0, 1], and compare it with Boltzmann factor, exp(-$\delta\varepsilon/k_B T$). If k is less than the factor, set $\vec{R}$ as $\vec{R}+\vec{\Xi}$ and return to next particle.

The specific surface energy $\Delta E_S$ change is calculated by (F.16), where $\zeta$ (J/kg m$^2$) is the specific surface energy of unit area, to form an interface of $CuFe_2O_4$ crystal, there are 8 Fe-O bonds (710.5eV) and 6 Cu-O bonds (936.7eV) [15] needed to be break per area of lattice cell. $\Delta S$ is the change of specific surface area. $m$ is the mass summation.

$$\Delta E_S = \zeta \cdot m\Delta S \qquad (F.16)$$

The $\Delta S$ is calculated by an approximation of area, which is the summation $S_0$ of area of the two parts $L_0$ & $L_1$ of aggregation and then minus a correction item S', by considering the surface composed of triangular faces.

$$\Delta S = (S_0 + 4\pi R^2 - S')/nM \qquad (F.17)$$

$S'$ is related with the structure formed by the joint of $L_1$. $n$ is the amount of particles, $M$ is the mass per particle. In Fig.2, by forming each equilateral triangle with tangent particles, the surface reduce $S_3$. By forming each regular tetrahedron with tangent particles, the surface reduce $S_4$.

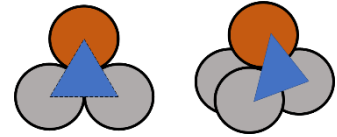

Fig.2 Approximation of surface area change

$$S_4 = 2S_3 = (2\pi - \sqrt{3})R^2 \qquad (F.18)$$

The number of formed equilateral triangle and regular tetrahedron can be calculated by the known coordinates of particles in $L_0$.

Consider a single sphere particle in $L_0$, its Lennard-Jones potential with $L_1$ is

$$V(r) = 4\epsilon\left[\left(\frac{\sigma}{r}\right)^{12} - \left(\frac{\sigma}{r}\right)^{6}\right] \quad (F.19)$$

Where $r$ is the distance between the two particles. $\epsilon$ and $\sigma$ are parameters. The total potential between $L_0$ and $L_1$ is

$$V = \sum_{i=1}^{n-1} V(r_i) \quad (F.20)$$

Based on the above formulation, the potential energy of a single nanoparticle can be denoted by relative cohesive energy $E_a/E_0$ [16]. $E_a$ is the cohesive energy per particle. $E_0$ is the total cohesive energy of the aggregation of particles. As the packing density of HCP and face center packing is same, the relation between $E_a/E_0$ and size of nanoparticles was calculated and shown in Fig.3.

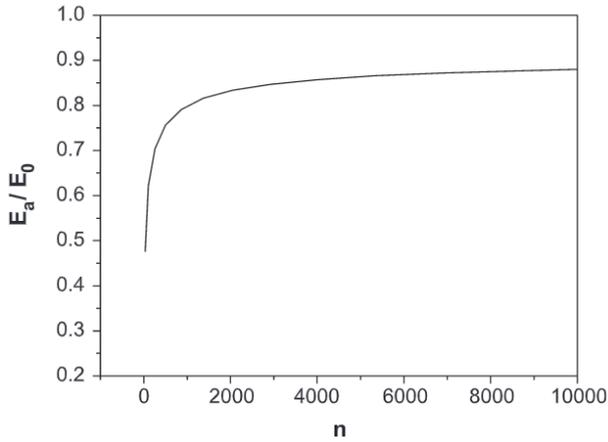

Fig.3 The particle size dependence of the cohesive energy of face-centered cubic nanoparticles [14].

Thus, the potential energy change during the j$^{th}$ particle with radius as $R$ aggregate on the initial particle can be expressed as (F.21), where $\varepsilon_0$ is the potential energy of the first two aggregated particles.

$$V_n = \varepsilon_0 \left(\left.\frac{E_a}{E_0}\right|_{n=R}\right)^j \quad (F.21)$$

The morphology of nano particles aggregation was evaluated by ratio $\omega$ of the module of vector of final particle to the summation of diameter of all particles. The ideal uniform aggregation should be a sphere, consider the volume conservation, there is following formulation

$$R' = 2R\sqrt[3]{n} \quad (F.22)$$

where $R'$ is the radius of the sphere aggregation assumed, $n$ is the amount of total particles calculated in the simulation. The closer $\|\vec{R} + \vec{\Xi}\|$ to $2nR$, the more ununiform morphology is. The closer $\|\vec{R} + \vec{\Xi}\|$ to $R'$, the more uniform the morphology is. Thus, $\omega$ can be denoted as (F.23), which should in [0, 1].

$$\omega = \frac{\|\vec{R} + \vec{\Xi}\| - R'}{2nR - R'} \quad (F.23)$$

## 2 Result and Discussion

### 2.1 Relation of Nucleation Rate with Temperature

**Table.2** Parameters of solvents used in simulation

| Solvent | $M_i$ g/mol | Density g/mL | Boiling point °C |
|---|---|---|---|
| $H_2O$ | 18.02 | 0.997 | 100 |
| $C_6H_{14}O_4$ | 150.17 | 1.1 | 285 |
| $C_8H_{18}O_5$ | 194.23 | 1.1 | 327 |
| $C_{18}H_{34}O_2$ | 282.47 | 0.895 | 360 |

The above four solvents were chosen to be calculated in this article, $H_2O$ is a reference. The curves of nucleation rate with temperature of the four solvents were shown in Fig.4. The calculation shows that $T_{min}$ is below the lowest temperature in experiments, $T_{max}$ is over 350°C, so the temperature range is from 50°C to boiling point of each solvent until 350°C.

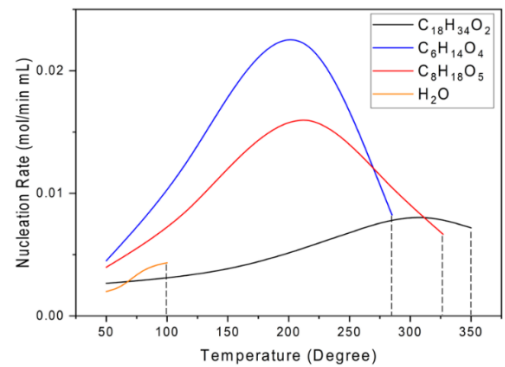

Fig.4 Modeling curve of nucleation rate of four solvents

As it was shown in Fig.4, the nucleation rate in water is pretty low when the temperature is below its boiling point at normal pressure, so that the hydrothermal synthesis is always done in high pressure containers. The nucleation rate of $C_{18}H_{34}O_2$ is low due to its high molar mass and viscosity. By setting the reaction temperature between 250°C to 300°C, the nucleation rate in $C_6H_{14}O_4$ and $C_8H_{18}O_5$ will pass the highest peak during the cooling process, it can form the most nucleus without boiling the solvent at normal pressure, which can reduce energy consumption and simplify equipment used in preparation.

**2.2 Relation of Nucleus Radius with Time**

The diffusion coefficients of iron and copper ions in water under temperature which below its boiling point at normal pressure are too low to be applied in the model. The total cooling time means the time caused of the total process from reaction temperature to 50°C, which affects the lapse rate of temperature, it is not the time since cooling beginning.

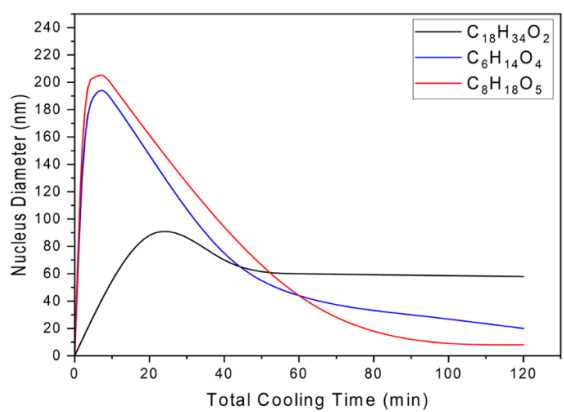

Fig.5 Modeling curve of nucleus radius

In the Fig.5, the increasing trend of $C_6H_{14}O_4$ and $C_8H_{18}O_5$ shorter than 10min, $C_{18}H_{34}O_2$ shorter than 30min, which is the fast cooling of the system, at the beginning if the system cools down to 50°C in an extreme short time, the nucleuses have no time to grow as the diffusion coefficients of ions were already reduced quickly to values too low to grow. When the cooling time was slightly extended but still short, there will not form a quantity of nucleuses, but the growth could only happen on these few nucleuses then make their diameter reach a high value. As the total cooling time was extended longer, there will be more and more nucleus formed during the time, then the diameter of each particle of different cooling time system decreases.

**2.3 Morphology Uniformity**

Assume the reaction temperature is 280°C, then was cooled to 50°C, the curve of $\omega$ with total cooling time was shown in Fig.6. The morphology uniformity of fast cooling calculated by the model is meaningless and was deleted from the figure. It can be seen from Fig.6, as the total cooling time extends, the morphology uniformity increased, which is because there are more time for particles to change locations until the low energy position was found [17].

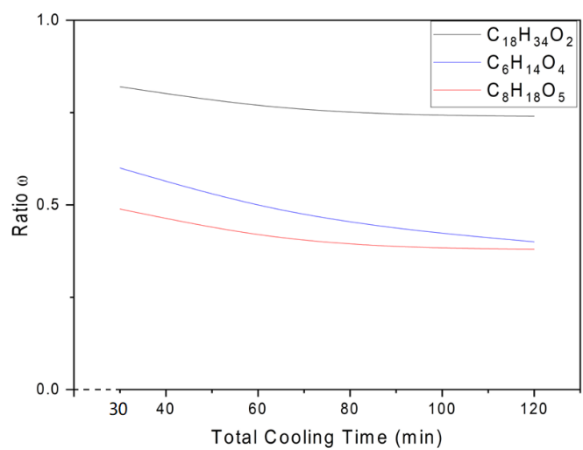

Fig.6 Modeling curve of morphology uniformity

## 3 Conclusion

By simulating the nucleation process of copper ferrite nano particles and calculated the average radius of particles in preparation systems, solvents as $C_6H_{14}O_4$, $C_8H_{18}O_5$ are the appropriate choices. Temperature between 250-300°C and cooling time over 1h is suitable for particles below 50nm. By applying Monte Carlo method to evaluate the morphology uniformity, cooling time is important to the aggregation of particles.

With the models built in this article, it will provide reference and choice of solvents, temperatures and cooling times when considering about optimization of the nano properties of copper ferrite during preparation.

**Acknowledgment**

The support from Bachelor program in Institute of Inorganic and Nonmetal Matter, HEU. Quite appreciate the care and support of W.B. Z. and PhD students in his group, CCME, PKU.